# An Integrated Crosscutting Concern Migration Strategy and its Application to JHoTDraw

Marius Marin, Leon Moonen and Arie van Deursen

**TUDelft**

**SERG**

# An Integrated Crosscutting Concern Migration Strategy and its Application to JHOTDRAW

**Marius Marin**
*Delft University of Technology*
*The Netherlands*
A.M.Marin@tudelft.nl

**Leon Moonen**
*Delft University of Technology*
*The Netherlands*
Leon.Moonen@computer.org

**Arie van Deursen**
*Delft Univ. of Technology & CWI*
*The Netherlands*
Arie.vanDeursen@tudelft.nl

## Abstract

*In this paper we propose a systematic strategy for migrating crosscutting concerns in existing object-oriented systems to aspect-based solutions. The proposed strategy consists of four steps: mining, exploration, documentation and refactoring of crosscutting concerns. We discuss in detail a new approach to aspect refactoring that is fully integrated with our strategy, and apply the whole strategy to an object-oriented system, namely the JHOTDRAW framework. The result of this migration is made available as an open-source project, which is the largest aspect refactoring available to date. We report on our experiences with conducting this case study and reflect on the success and challenges of the migration process, as well as on the feasibility of automatic aspect refactoring.*

## 1. Introduction

The tangling and scattering that results from implementing crosscutting concerns in a software system using traditional object-oriented programming is a known challenge to program comprehension and software evolution. One approach to mitigate these issues is to migrate the system to aspect-oriented programming (AOP) and transform the crosscutting concerns into aspects, a process known as *aspect refactoring*.

Despite significant research efforts on various parts of the refactoring of crosscutting concerns from existing systems, to date there exists no compelling show-case for such a complete migration. One of the main causes for this void is the fact that there is no clearly defined, coherent migration strategy detailing the steps to be taken to perform this process.

Successful migration requires a strategy comprising steps like identification of the concerns (i.e., aspect mining), description of the concerns to be refactored, and consistent refactoring solutions to be applied. Moreover, such a strategy requires *integrated* migration steps, so that aspect mining results, for example, can be consistently mapped onto concerns in code, and further refactored by general aspect solutions. The present state of the art prevents developers and practitioners from experimenting with a complete migration process and assessing the benefits of migrating to AOP.

In this paper, we propose such an integrated strategy for migrating crosscutting concerns to aspects, which consists of four main steps: (1) idiom-driven identification of crosscutting concerns in an existing system (aspect mining); (2) exploration of (the context of) the concerns identified in the previous step; (3) query-based modeling and documentation of crosscutting concerns in the system; (4) template-based refactoring of the object-oriented idioms into AOP solutions.

Our strategy builds upon the classification and decomposition of crosscutting concerns in so-called *crosscutting concern sorts* that we proposed earlier [11, 13]. Each sort describes the typical implementation idiom and relation of crosscutting concerns. Sorts act as glue between the successive steps of the migration: The mining step in our strategy uses the sort-specific idioms to define search-goals for identifying crosscutting concerns that belong to a specific sort (i.e., *sort instances*). To support the exploration and documentation steps, we have formalized the concern sorts using queries over source code and implemented these in a tool for browsing and modeling crosscutting concerns [14].

While the first three steps of our approach have been covered in our earlier work, this paper focuses on the fourth step and its connection with the three preceding steps. In particular, we define template solutions for the aspect refactoring of our sorts (to AspectJ). Furthermore, we describe a case study in which we apply the whole migration strategy to JHOTDRAW,[1] an object-oriented application used in other aspect mining and refactoring studies as well [10, 2, 12, 1]. The results of our migration are available under version control as an open-source project on sourceforge called AJHOTDRAW, which is also the largest aspect refactoring publicly available to date that we are aware of.

The remainder of the paper is organized as follows. In next section, we recall the notion of crosscutting concerns sorts. We describe the migration strategy and elaborate on the first three steps in Section 3. The sort-based aspect refactoring approach that we introduce for the fourth step is presented in Section 4. Section 5 covers our experiences with migrating crosscutting concerns in JHOTDRAW to aspect solutions. Section 6 discusses the results and outlines a number of lessons learned. We conclude with an overview of related work and recommendations for future research.

---

[1] `http://jhotdraw.org`

## 2. Crosscutting concern sorts

A systematic migration strategy requires a consistent way to address crosscutting concerns in source code. To this end, we distinguish a number of *atomic* crosscutting concerns (i.e., concerns that cannot be split into smaller, still meaningful concerns) that share properties like their implementation idioms and relations. We group concerns that share such properties in categories called *crosscutting concern sorts* [11]. These sorts can be used on their own, but can also be composed to construct more complex crosscutting designs, for example, the *Observer* pattern, often used as a typical example of crosscuttingness.

The first two columns of Table 1 describe the identified sorts and show several examples of instances (the other columns will be introduced in later sections). *Consistent behavior*, for instance, groups concerns whose implementation consists of scattered calls to a specific method implementing the crosscutting functionality. Instances of this sort include, for example, a logging concern, a simple authentication or authorization concern implemented as a call to a method checking credentials, or a mechanism for updating observers using calls to a notification method.

Similarly, the idiom for implementation of secondary roles, common in design patterns like *Observer* or *Visitor*, as well as in mechanisms for persistence, is described by the *Role superimposition* sort.

Composite crosscutting designs exhibit multiple sort instances in their implementation: the aforementioned *Observer* pattern, for example, comprises two instances of *Role superimposition*, for the Subject and the Observer role respectively. Furthermore, it comprises instances of *Consistent behavior*, like the concern for notification of observers, or the one for observers registration. Instances of our sorts are therefore *building blocks* for modeling and describing crosscutting functionality.

## 3. An integrated migration strategy

In this section, we define an integrated strategy for migrating crosscutting concerns in existing systems to aspect-based solutions. The strategy consists of four steps:

**Step 1.** Idiom-driven crosscutting concern identification (also known as *aspect mining*).

**Step 2.** Concern exploration.

**Step 3.** Query-based concern modeling and documentation.

**Step 4.** Sort-based aspect refactoring.

The remainder of this section discusses the first three steps in more detail and the next section presents the fourth step. We show how the steps are integrated via crosscutting concern sorts using examples from our JHotDraw to AJHotDraw migration experience.

### 3.1. Aspect mining

In our earlier work we have proposed and implemented an idiom-driven approach to aspect mining based on crosscutting concern sorts [12]. The approach supports the design of aspect mining techniques that target instances of a specific sort by searching for the sort's implementation idiom.

The third column in Table 1 shows the implementation idioms associated with each of the sorts. Consider for example the commands in a drawing application, like JHotDraw, that carry out tasks in response to user actions. Each command concludes its execution with a call to the `checkDamage` method in the drawing view, which updates the view with changes triggered by the command. The notification concern is an instance of *Consistent behavior* whose implementation idiom is invocation of a specific method from a (large) set of methods. Aspect mining techniques such as Fan-in analysis [10] or Grouped calls analysis [12] exploit idioms such as this one in their search process.

We have implemented the two mining techniques mentioned above and an additional technique that targets instances of *Redirection layer* in our aspect mining tool FINT[2] [10, 12]. The results of applying FINT to JHotDraw are the starting point of our migration case study.

Like the notification mechanism above, we have found the *Consistent behavior* idiom in multiple concerns implementing support for commands and undo operations. Examples include consistently checking the reference to the active view before execution of each command, consistent initialization of Command objects by means of super calls, or consistent checks implemented by all actions to undo a command. Our search for idioms of the *Redirection layer* pointed us to wrapper objects for undo-able commands: methods in the wrapper delegate calls to their wrapped command object.

### 3.2. Concern exploration

Aspect mining often does not yield complete crosscutting concern instances, but just concern *seeds*: (possibly incomplete) sets of program elements that belong to a particular crosscutting concern.

The second step of our strategy, concern exploration, aims at expanding mining results (i.e., concern seeds) to the complete implementation of the associated concerns. In this step, we start from the discovered seeds and use the specific relation of the sort for the seed's concern to identify all the participants in the concern implementation.

In our *Consistent behavior* example, this means looking at all call relations directed to the method `checkDamage` (or another method, depending on the particular concern targeted). As it turns out, not all of the 28 calls to this method that we found are part of the concern of interest, but around two-thirds of them, namely those from *Command* classes. Similarly, the *Grouped calls* mining technique, which applies

---

[2] Available from http://swerl.tudelft.nl/view/AMR/FINT

| Sort and Intent | Examples | Idiom | Template aspect solution |
|---|---|---|---|
| *(Method) Consistent Behavior (CB)*: A set of methods consistently invoke a specific action as a step in their execution. | Logging of exception events in system; Wrapping business service exceptions and re-throwing them as new exception type [10]; Notification of Figure change events. | Method invocations from set of methods. | Pointcut and advice mechanisms.<br>`around(..) : callersContext(..){`<br>`    invokeCB(..); //before`<br>`    proceed();`<br>`    // or after: invokeCB(..);`<br>`}` |
| *Redirection Layer (RL)*: A type acts as a front-end interface having its methods responsible for receiving calls and redirecting them to dedicated methods of a specific reference, optionally executing additional functionality. | Border decorations for Figure elements (Decorator pattern); Command wrapper for undo support. | Redirector type whose methods consistently forward calls to pair methods in receiver. | Pointcut and around advice to replace each redirection.<br>`around(..) : call Receiver.m(..) &&`<br>`    filteredCallers(..) {`<br>`    addBehavior1();`<br>`    proceed(..); //redirection`<br>`    addBehavior2();`<br>`}` |
| *Expose Context (EC)*: Context Passing: Methods in a call chain consistently use parameter(s) to propagate context information along the chain. | Transaction management [8]; Credentials passing for authorization; Progress monitor for long-running operations. | Method in chain passes parameter as argument to callee. | Pointcut and advice, where the point cut collects the context to be passed - Wormhole [8]<br>`around(<caller context>, <callee context>):`<br>`    cflow(callerSpace(<caller context>)) &&`<br>`    calleeSpace(<callee context>){`<br>`    // ... advice body`<br>`}` |
| *Role Superimposition (RSI)*: Types extend their core functionality through the implementation of a secondary role. | Figure elements observed by views for changes (Subject role); Visitable elements (Visitor pattern); Storable figures (Persistence) [10]. | Set of types (declare and) implement member roles (which are possibly declared by a distinct interface). | Introduction mechanisms.<br>`declare parents :`<br>`    Type implements SecondaryRole;`<br>`Modifiers Type Type.roleField;`<br>`Modifiers Type Type.roleMethod(..){`<br>`    ...//original implementation`<br>`};` |
| *Support Classes for Role Superimposition (SC)*: Types implement secondary roles by enclosing nested support classes. The nesting enforces (and explicates) the relation between the enclosing and the support class. | Undo support for Command elements; Event dispatcher for observers' notification. | Set of types (in hierarchy) implement Role using nested classes. | The desired solution, introduction for nested classes, is not supported by AspectJ. Our solution is to move the support classes to the aspect. |
| *Exception Propagation (EP)*: methods in call chain consistently (re-)throw exceptions from their callees in the absence of an appropriate answer. | IOException thrown if Figure elements recovery fails; Checked SQLException thrown from methods in the JDBC API. | Method in call chain re-throws exception to caller. | Softening exceptions mechanisms.<br>`declare soft : ExceptionType :`<br>`    (call(* rootException(..)`<br>`        throws ExceptionType));`<br>Capture SoftException at top of the call chain. |

**Table 1. Crosscutting concern sorts.**

a more conservative search, covers only partially the set of calls participating in the concern.

A number of tools provide (partial) support for exploring seeds and expanding them to full concerns, and for querying source code for concern sort relations: FINT, the Eclipse IDE, the Concern Manipulation Environment (CME) [18], FEAT [16], JQuery [7], CodeQuest [4], or SOQUET [14]. The same tools can be used to further understand the context enclosing the discovered crosscutting concern. At this step, we can see, for example, how the identified sort instances in command and undo support relate to each other: commands that can be undone enclose a specialized *UndoActivity* class that knows how to revert the effects of the command's execution. Two of our mined sort instances cover the key methods of the two classes: the `execute` method in a command, and the `undo` one in the enclosed undo activity.

### 3.3. Concern modeling and documentation

Most approaches to concern modeling and their tool support do not enforce consistency across the representation of crosscutting concerns. The decision of what is crosscutting in a system, and how to best represent that, lies with the user of such concern modeling tools. Such a concern model can contain ad-hoc collections of program elements, like methods and classes, that participate in a concern's implementation.

However, to ensure generally applicable solutions for concern migration, we need a coherent way to describe similar concerns and their common properties. To this end, we have defined queries for each of our crosscutting concern sorts which search for the sort's specific relation between source code elements. For more information on these sort queries, we refer to our earlier work [13], which formalizes these queries using relation calculus over source models extracted from the system's source code.

We have implemented support for this third migration step in our concern modeling tool SoQueT[3] [14]. Figure 1 shows two of the main views of the tool. The *Concern model* view allows us to organize concerns hierarchically, with sort instances and their associated queries as leaf-elements and composite concerns describing more complex crosscutting designs as parents. The user can select a sort instance in the concern model and execute its query; The results of the query are displayed in the *Search (Sorts Result)* view, from where they can be navigated to their source code implementation. To add a new sort instance to the model, the user launches the dialog providing the query templates for each sort, and parameterizes the query for a given crosscutting concern. For example, to document our *Consistent behavior* instance for notification of views, we use the knowledge gained at the previous steps and search for all the calls to the `checkDamage` method from methods in the *Command* hierarchy. The method and the hierarchy are our input parameters to the query. The instances can then be added to the model from the results view.

A part of the concern model built to document concerns in JHOTDRAW is shown in Figure 1. The model is available for download at the same web-site as the tool and covers over 100 sort instances.[3] In Section 5, we use this documentation to guide our refactoring and configure the aspect solutions.

## 4. Aspect refactoring

We employ a sort-based, idiom-driven approach to aspect refactoring that allows for consistent integration with the previous steps of our migration strategy. Furthermore, we define template aspect solutions for each of our concern sorts that we can instantiate to refactor an occurrence of that sort. Like the previous steps, the refactoring approach addresses crosscuttingness at the level of atomic concerns, which provides the optimal trade-off between complexity of the refactoring and comprehensibility of the refactored element.

The template aspect refactorings for each sort are summarized in the last column of Table 1. A solution basically consist of one aspect language mechanism. At the moment, however, some sorts do not have an equivalent mechanism in AspectJ (or any other aspect language existing at this moment). Support classes, for example, cannot be introduced similarly to role members, although, as we shall see in Section 6, this would be a desired refactoring.

To refactor a sort instance, we start from its query-based documentation (in SOQUET). The query points us to the elements participating in the concern, which we use to configure the template aspect solution. For example, the query for a *Consistent behavior* instance indicates the callers to be captured by a pointcut definition (the source context) and the action to be introduced by the advice (the target context). Other configurable elements, such as the type of advice to introduce the crosscutting call (e.g., before, after, after throwing, etc.), are decided at the refactoring time.

The solution described in Table 1 for the *Redirection layer* sort is a common approach to refactoring implementations of the *Decorator* pattern [5, 9]. This consists of replacing the redirector class by an aspect that intercepts (relevant) calls to the methods receiving the redirection, and then adds the redirector's functionality by means of an advice.

The aspect solution for *Expose context* instances is discussed by Laddad as the *Wormhole* pattern [8]: the extra parameter used to pass context is replaced by using a pointcut to obtain the context from the caller and an advice that makes the context available to the caller's control flow.

Solutions for static crosscutting, like *introduction* and *declare soft* mechanisms in AspectJ, apply to two of the sorts in the list, *Role Superimposition* and *Exception propagation* respectively. The elements to instantiate these aspect templates are again available through the sort-based documentation of the concerns: they indicate the members of a type's secondary role to be moved to and then introduced from an aspect, or the checked exception to be turned into an AspectJ soft exception. Soft exceptions, unlike checked ones, do not need to be caught or re-thrown. This allows us to remove the *throws* clauses from the (transitive) callers of the method initiating the exception propagation. The method at the top of the call chain that deals with the exception has now to catch the soft exception that wraps the original checked one. The top method assumes knowledge of the wrapped exception that it has to extract and cast. The code to handle the (cast) exception requires no modifications.

## 5. Aspect Refactoring of JHOTDRAW

We have used the sort-based migration strategy to refactor a number of crosscutting concerns in JHOTDRAW towards an aspect-oriented solution. Based on these experiments, we would like to obtain answers to the following questions:

1. Are the template aspect solutions proposed in Section 4 applicable in practice?
2. What are the risks and benefits of adopting refactoring strategy that is sort-based?
3. What level of automation of all four steps and the fourth refactoring step in particular is feasible?
4. Do the refactorings carried out lead to a better separation of concerns?

In the present section, we report our observations and experiences regarding the migration of specific crosscutting concerns towards aspects in JHOTDRAW. In the next section, we return to our questions, and try to formulate answers to them based on the findings presented here.

---

[3] Available from http://swerl.tudelft.nl/view/AMR/SoQueT

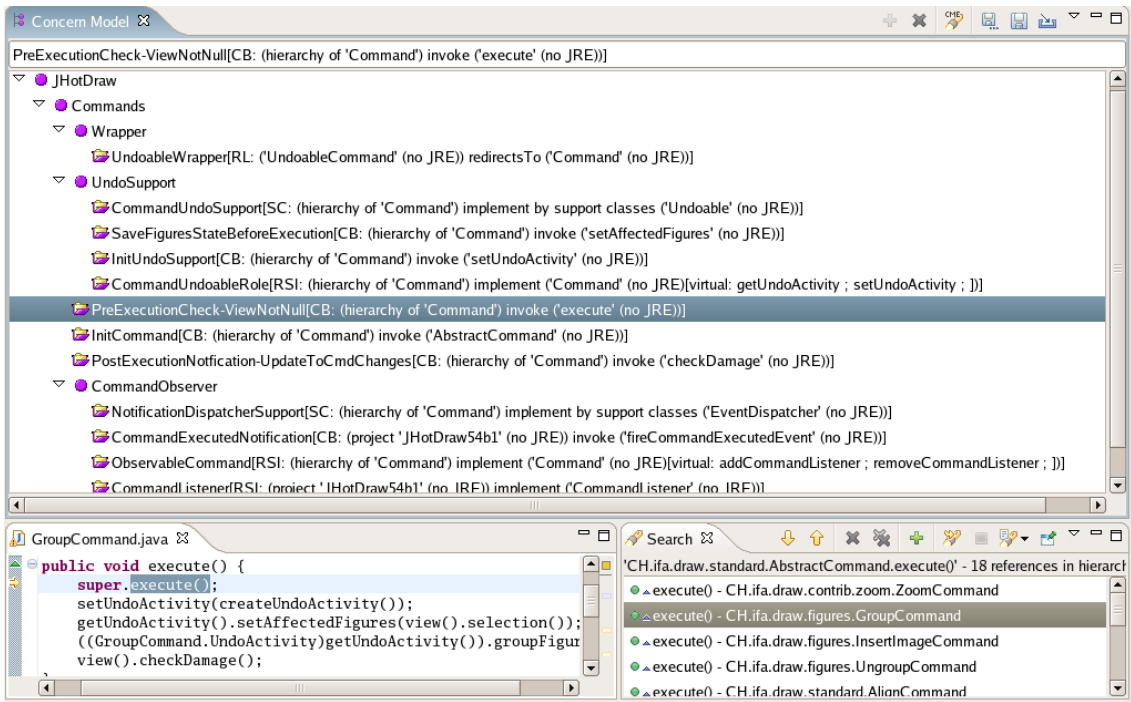

**Figure 1.** SOQUET **documentation of the concerns for Command support in** JHOTDRAW**.**

## 5.1. AJHOTDRAW

We share the refactored version of JHOTDRAW as an open-source project on sourceforge[4]: AJHOTDRAW is, to our knowledge, the largest migration to aspects available to date. A transparent, gradual migration process is important for building confidence in the aspect-oriented solution. Therefore, our refactorings aim at maintaining the conceptual integrity and stay close to the original design. In addition, by publishing the refactoring steps in a versioned repository, we provide insight in the migration process and enable traceability, making the refactored system easier to understand.

Our next discussion focuses on the refactoring of sort instances contained in the implementation of the command and undo functionality, which we also used in Section 3 to explain the first three steps of the approach. We use the organization of concerns in the concern model initiating the refactoring to design the package and type structure of our aspect solutions. The solutions discussed below have been integrated with the source code available on the public repository.

## 5.2. Consistent behavior in Command

JHOTDRAW makes use of the *Command* design pattern in order to separate the user interface from the underlying model, and to support such features as undoing and redoing user commands. Each command has to realize the *Command* interface, for which a default implementation is provided in the *AbstractCommand* class. The key method is *execute*, which takes care of actually carrying out the command (such as pasting text, duplicating a figure, inserting an image, etc.).

A typical implementation of a command is highly crosscutting, with the *Command* top interface defining three different roles: besides their core functionality, commands are undo-able as well as observable elements. The support for the secondary roles counts for half of the *Command*'s members. Similarly, the execute method in a typical concrete command implements multiple concerns.

Each execute method should start with a consistency check verifying that the underlying "view" exists. Therefore, each concrete implementation of execute starts with a call

```
public class AbstractCommand implements Command {
  ...
  public void execute() {
    if (view() == null) {
      throw new JHotDrawRuntimeException(
        "execute should NOT be getting called when" +
        "view() == null");
} } }

public class PasteCommand extends AbstractCommand {
  ...
  public void execute() {
    super.execute();
    ...
} }
```

**Figure 2. Consistent check - super method idiom.**

---
[4] http://sourceforge.net/projects/ajhotdraw/

```
pointcut cmdExecute(AbstractCommand aCommand) :
  this(aCommand)
    && execution(void AbstractCommand+.execute())
    && !within(*..DrawApplication.*);

before(AbstractCommand aCommand) : cmdExecute(aCommand) {
  if (aCommand.view() == null) {
    throw new JHotDrawRuntimeException("...");
  } }
```

**Figure 3. Enforcing consistency using advice.**

to the `execute` implementation in the superclass, which is always the one from the *AbstractCommand*. This is illustrated in Figure 2.

We apply a *Consistent behavior* refactoring template from the last column in Table 1 using a pointcut capturing all `execute` methods, and putting the check itself in the advice. Observe that mimicking the implementation where the check is in a super method is not possible in AspectJ: super methods cannot be accessed when advising a method. The resulting solution is shown in Figure 3.

The only surprise in this code may be the `within` clause in the pointcut. In the exploration step, we learned that *anonymous* subclasses of *AbstractCommand* do not implement the consistency check. Such classes are used for simple commands like printing, saving, and exiting the application. Since AspectJ does not provide a direct way to exclude anonymous classes in a pointcut, we used the `within` operator to exclude executions occurring in the context of the top level object creating the full user interface. One can also argue that the anonymous classes should include this check (in which case the exclusion can be omitted from the pointcut), but, as stated before, we focus on keeping the behavior as it was, not on modifying it.

Besides the separation of the consistency check from the core logic of the commands, another benefit of the aspect approach is that consistency checks cannot be forgotten. This is illustrated by a number of the anonymous classes, but also by one non-anonymous command,[5] which does not extend the *AbstractCommand* default implementation. Consequently, it cannot reuse the consistency check using a supercall. Inspection of the `execute` implementation, however, clearly shows that the code exits with a null pointer exception in case the check fails. This suggests that the aspect that we are looking for should implement the check not only for the *AbstractCommand* class, but for all the *Command* implementations.

### 5.3. Undo Functionality

Support for "undo" functionality was added in JHOTDRAW version 5.4. As can be imagined, it is a concern that cuts across many different classes. More than 30 elements of the JHOTDRAW framework, comprising *commands*, *tools* and *handles*, have associated undo constructs to revert the changes spawned by their underlying activities. The *commands* group is the largest in terms of defined undo activities.

The participants of the "Undo" functionality have the following responsibilities:

- Each command is associated with one *undo activity*, whose method *undo* can be invoked to revert the command. The undo activity is implemented in a nested class of the command, which is instantiated using a factory method called `createUndoActivity`.
- Prior to the execution of the command's core logic, the command saves a reference to its associated undo activity, by calling a dedicated setter method.
- The primary abstraction in the undo activity is the list of affected figures: when the command's `execute` method is invoked, the relevant state of the affected figures is stored in the undo activity.
- Undo activities are maintained on a stack by the undo manager.

#### 5.3.1. Support classes for role superimposition

The refactoring that we propose for Undo consists of associating a dedicated undo-aspect to each undo-able command. The aspect implements the entire undo functionality for the given command, while the associated command class remains oblivious to its secondary (undo) concern.

We use naming conventions to relate the aspect to its supported command class. In a successive step, we refactor each of the sort instances in the undo support. The command's nested *UndoActivity* class belongs to a *Support classes* instance. In the absence of introduction mechanisms for nested classes in AspectJ, our aspect solution consists of moving the *UndoActivity* class into the aspect.

The factory methods for the undo activities (`createUndoActivity()`), as well as the members for managing the reference to the command's undo activity belong to an instance of *Role superimposition*. The role members move to the aspect, from where they are introduced back into the associated command classes using inter-type declarations. The design, however, suffers modifications as the visibility of the undo factory methods has been altered: ASPECTJ cannot be used to introduce the required factory method as *protected*.

#### 5.3.2. Consistent behavior

The invocations in the `execute` method that are responsible for setting up the undo activity implement *Consistent behavior* concerns: the calls are taken out of the `execute` method, and woven into it by means of advice. In some cases the corresponding pointcut simply needs to capture all `execute` method calls. However, in other cases the pointcut is more complex, depending on the way the undo code is mixed with the regular code.

---
[5]Namely, the *UndoableCommand*.

```
public class PasteCommand extends FigureTransferCommand {
  public void execute() {
    ...
    FigureSelection selection = (FigureSelection)
        Clipboard.getClipboard().getContents();
    if (selection != null) {
        setUndoActivity(createUndoActivity());
        ... //core command logic and other undo setup
        FigureEnumeration fe = insertFigures(...);
        getUndoActivity().setAffectedFigures(fe);
        ...
} } }
```

**Figure 4. The original PasteCommand class.**

As an example to illustrate that automating such refactorings is not at all straightforward, consider the paste-command, whose `execute` method consists of retrieving the selected figures from the clipboard, inserting them into the current view, and clearing the clipboard. All this is done in a single method, using local variables and if-then-else statements to deal with situations like pasting from an empty clipboard. The undo aspect will require the same conditional logic, and access to the same data in the same order. The following alternatives are possible for aspect refactoring:

- if all getters are side effect free, an approach is to setup the undo activity in a simple before advice. In JHOTDRAW, however, this is not the case, for example because of figure enumerators that have an internal state.
- an alternative is to intercept relevant getters, keep track of the data locally in the advice as well, and inject advice after all data has been collected. This is the approach we follow, but some of the pointcuts are somewhat artificial. Figure 5 shows such a pointcut in the undo aspect for the *PasteCommand*, refactored from Figure 4. The *clipboardGetContents()* pointcut captures the call that sets the reference to be checked by both the command's core logic and the undo functionality in the aspect.
- The last possibility is to refactor the long `execute` method into smaller steps using non-private methods. The extra method calls can be intercepted allowing smooth extension with setting up the undo activity, at the cost of creating a larger interface and breaking encapsulation. Moreover, we would still introduce artificial pointcuts, as our intention is to enhance the behavior of the `execute` method, and not of various steps created for supporting advice introduction.

#### 5.3.3. Redirection layer

The design of undo in JHOTDRAW uses wrapper objects to associate undo-able commands to menu items and buttons in the user interface (UI). The wrappers share their top level interface with regular commands, so they can connect to UI elements and receive user actions. While most commands are undo-able and wrapped by an *UndoableCommand* object, there are a few exceptions, such as, *CopyCommand*.

```
public aspect PasteCommandUndo {
  //store the Clipboard's contents - common condition
  FigureSelection selection;

  pointcut clipboardGetContents() :
      call(Object Clipboard.getContents()) &&
      withincode(void PasteCommand.execute());

  after() returning(Object select):clipboardGetContents(){
        selection = (FigureSelection)select;
  }
  ...

  pointcut executePasteCommand(PasteCommand cmd) :
      this(cmd) && execution(void PasteCommand.execute());

  // Execute undo setup
  void after(PasteCommand cmd):executePasteCommand(cmd) {
      // the same condition as in the advised method
      if(selection != null) {
          cmd.setUndoActivity(cmd.createUndoActivity());
          ...
          cmd.getUndoActivity().setAffectedFigures(...);
} } }
```

**Figure 5. The undo aspect for PasteCommand.**

Wrappers are instances of *Redirection layer*. The refactoring of such instances raises several important issues: first, we need to identify those commands that are wrapped by an *UndoableCommand* object and accessed through this object; second, we need to check if all clients of a command access its functionality via the wrapper. Only those calls from command clients that are received by a wrapper in the original implementation need to be captured by the aspect solution to attach the wrapper's functionality by means of advice.

Further complications that limit feasibility of automated refactoring have to do with the multiple roles in *UndoableCommand*: since the aspect solution completely replaces the wrapper class, this means that introduction of roles is no longer possible. Some of the original roles in the system are implemented by the wrapper only to comply with the top interface of the wrapped element and add no specific functionality, such as the *Observable* role of *Command*s. The aspect solution can safely omit these roles. For other roles however, this is not desired and refactoring requires customized redirector solutions.

## 6. Discussion

**Applicability in practice** The proposed template aspect solutions proved suitable for refactoring concrete sort instances in the JHOTDRAW case and for separating the crosscutting code from the core system. However, the difficulty of implementing the aspect solution and the quality of the result will vary from case to case. One of the issues is pointcut definitions: Ideally, we would like to use pointcut definitions that describe a set of elements by formalizing a common property instead of a brittle enumeration of the elements in the set. In practice, such definitions will not always be feasible, either because of limitations in the aspect language, or due to irregularities in the code under investigation.

| Sort | Limitations and risks |
|---|---|
| Consistent Behavior | Advice constructs in a privileged aspect can break encapsulation; High degree of tangling might prevent (automatic) refactoring; Anonymous classes cannot be referred to consistently, preventing generic pointcuts; Calls to super class functionality cannot be migrated into advice; Modular reasoning affected by need to keep track of data set in the advised method; Check required that omissions are not on purpose; Sophisticated pointcuts needed to intercept all relevant state modifications in the advised methods; Check required that precedence does not change due to new advice; |
| Redirection layer | The repetitive logic of redirection for the redirector's methods is not eliminated – the aspect solution addresses the redirection at method level and not at type level; New redirector methods are not (automatically) covered by the solution; The aspect solution is not dynamic (dynamic reordering of redirectors) [5]; The aspect solution replaces the redirector (wrapper) and hence changes the public interface of the application to test against; The calls (to the receiver) to be advised for redirection need to be detected; |
| Role superimposition | Visibility affected since protected (/non-public) methods cannot be introduced. |
| Support classes for role superimposition | Not supported; Nesting the support class in the aspect breaks dependencies (thus forcing the enclosing class to make more of its interface public) and weakens the relation with the enclosing class; |
| Exception propagation | Type of thrown exception is lost; Refactoring *throws* clauses in inheritance hierarchy. |

**Table 2. Risks and possible limitations of the aspect solution.**

Desired functionality included for example a pointcut to capture calls from "all *Command* classes, except all anonymous classes", or a pointcut for "all objects interested in command events". Irregularity in the code might require that for certain methods the advice executes only if a specific condition holds. This is the case for a few commands in JHOTDRAW that send notifications of their execution only if the clipboard's content is not empty. In such a situation, one has to make a trade-off between a generic pointcut definition that captures all commands, but ignores the particular condition, and a definition that enumerates all appropriate elements. The former solution would execute the code in the advice in spite of its void effect; however, the latter pointcut definition needs to be updated (manually) for every new element added to the set of interest (i.e., every new command).

Similar observations can be made about the definition of advices: sometimes we need to modify the original control flow of a method-to-be-migrated in order to introduce an action to it by means of advice. Although the refactoring may have no effect on the observable behavior of the method, the original flow could be more natural or comprehensible.

**Benefits and risks** In comparison with refactoring approaches proposed by others, our sort-based migration strategy gives a clear definition of the input required for refactoring (i.e., an atomic concern) and describes it consistently using queries. This allows for the definition of reusable solutions and improves comprehension of refactoring by addressing meaningful concerns instead of code fragments [1, 15]. Moreover, the concern queries allow us to describe the context cut across by a concern, and hence the concern's intent. This gives a better insight into the concern and its aspect solution than the simple enumeration of joinpoints common with most previous refactoring approaches. We believe that a clearly specified input for a refactoring solution is a necessary condition for ensuring consistent migration of concerns.

Among the main risks of refactoring, we identify the high level of coupling and complex dependencies between the base code and the crosscutting concern. We anticipate that any non-trivial aspect refactoring will require object-oriented refactorings, before the crosscutting concern can be taken out of the available system.

The issue with coupling is that, before migration, concern code can freely access certain parts of the core code that may have limited visibility after the migration. Possible risks in such a case are weakening the visibility restrictions of those members or violating encapsulation by declaring the aspect *privileged*. Other risks include code duplication in advice and the advised method or definition of artificial pointcuts to capture return values of calls from the advised method; this could be the case when some control logic is required by both aspect and the advised method.

We encountered several complex dependencies while refactoring instances of *Exception propagation* in JHOTDRAW. One example is the propagation of the *IOException* rooted in the set of methods to read drawings from file. The methods in the call chain re-throwing the exception override other methods, whose declared thrown exceptions might only serve for compliance with the method to be refactored. In this case, we also need to address their *throws* clause within our refactoring. Moreover, the overriding elements of a method in the chain that throw the same exception need to be refactored too, as their exception declaration is no longer allowed.

Table 2 summarizes the above risks and limitations in refactoring to aspects. Note that many of these limitations are independent of the strategy employed for refactoring. In spite of that, we are not aware of other papers in the area of refactoring to aspects that discuss these limitations.

**Automation** The refactoring step in our strategy currently has the least automation of all steps in our approach. However, the other (tool supported) steps give us many of the elements needed for refactoring, such as the crosscutting element and the context it cuts across which are captured in the concern documentation as repeatable queries. Moreover, the description of these elements by the sort-queries is similar to the definition of pointcuts for a possible aspect solution.

We believe that the case study presented in this paper is a required step before setting out to design (automated) aspect refactoring tooling. The study gives us insight into the complexity of each refactoring and the trade-offs to be made. The challenges and limitations discussed in the previous sections also indicate that completely automated aspect refactoring is unfeasible in any practical situation, since the process requires a significant level of interaction with the users to guide the system through the right decisions.

A particularly challenging automatic refactoring would be the one for *Redirection layer* instances: the original, dynamic solution uses a common interface for both redirectors and potential receivers. This interface hides the identity of the object for which a call is made; However, the refactoring of redirectors requires to know which calls are meant for a redirector and so need to be attached an advice introducing the functionality of the refactored redirector.

**Separation of concerns**   Our case study had a satisfactory outcome in achieving a better separation and modularization of concerns in the targeted application. As we were able to notice, the crosscutting code is an important part of the refactored elements, in some cases, such as the *Command* elements, over 50%. We appreciate that the core code is easier to understand in the absence of the migrated crosscutting concerns. To understand the aspect code, on the other hand, one typically also needs to understand the base code that it advises. This is exaggerated further by (high) coupling between the aspect and the base code, like for aspects that intercept calls from advised methods to reuse the values returned by such calls.

While refactored, crosscutting-free code is easier to comprehend, modifications to such code would still require awareness of the advice that applies to it. For instance, aspects might assume a certain order of the calls from an advised method, which has to be preserved to correctly introduce additional behavior.

Keeping track of the order of different advice in an aspect solution and preventing accidental changes might prove difficult, particularly when the number of aspects increases. The support from present development environments would not provide much insight into violations of such ordering, or into the ordering itself. This becomes more of an issue when the order is set using name-based wildcards, and new aspects match an existing rule for aspect precedence that should not apply to them. A similar situation might occur when changing an aspect solution that is already covered by a precedence rule, and the changes would not be compliant with that rule. Changing the position of an advice definition in an aspect could also modify precedence, if multiple advices in the aspect apply to the same joinpoints. Unspecified precedence could also lead to interference between new advices introduced by refactoring and existing ones [17]. Automatic refactoring needs to be aware of these issues.

Some concerns might be crosscutting for advices, similarly to the way they are crosscutting for methods. For instance, the re-use of specialized enumerations in JHOT-DRAW requires to reset them after each iteration. Such enumerations are used by some advices in the aspect solutions. Applying aspect solutions to aspects might prove challenging for both tool support and comprehensibility.

## 7. Related work

While each step in the migration of crosscutting concerns has been addressed by related research, we are not aware of an integrated strategy like the one proposed in this paper.

The present approaches to aspect refactoring can generally be distinguished by their granularity. Laddad's set of refactorings cover both low level ones, such as *extract method calls into aspects* or *extract interface implementation*, as well as more complex refactorings, like design patterns, transactions management, or business rules [8]. Although the latter subset typically involves multiple concerns to be refactored, there is no categorization of these concerns or refactorings.

Hannemann et al. propose an approach to the aspect refactoring of design patterns based on a library of abstract roles [6, 5]. The role-based refactoring requires one to map a pattern's implementation onto the predefined roles describing the pattern, and then applies a set of instructions to refactor the implementation to aspects. The approach is a step further towards generic, abstract solutions to typical problems that involve crosscutting functionality. However, as we have already seen, these patterns typically have a complex (and variable) structure in source code, which exhibits multiple (atomic) crosscutting concerns. The refactoring of a whole pattern in one step might prevent the comprehension of the concerns involved. Moreover, our experience suggests that pattern implementations can vary significantly from a standard description and one-step refactoring could be hampered by complex dependencies. We cannot make a full assessment of this approach as the implementation and the experimental results are not available, but we believe that all the limitations discussed in this paper would equally apply to it.

Finer-grained refactorings have been proposed in the form of code transformations catalogs [15] and AspectJ laws [3]. These transformations can occur as steps in the aspect refactoring of an (atomic) crosscutting concern, but remain oblivious to the refactored concern. They describe the mechanics of migrating Java specific units to AspectJ ones (e.g., *Extract Fragment into Advice, Move Method/Field from Class to Inter-type*). Such small step transformations might benefit the implementation of automatic refactorings by preventing complex dependencies and ensuring behavior preservation as discussed by Cole and Borba [3]. However, more effort is required to assess their general applicability: for example, the case-study used for the refactoring in [15], is an *Observer* pattern implemented in a demonstrative application, which lacks the complexity of a real system like JHOTDRAW.

In comparison to the work on fine-grained refactorings, the sort-based approach presented in this paper emphasizes concerns and identifies common properties at a consistent granularity level. This allows us to design a complete migration strategy, where the refactoring is integrated with steps for concern identification and comprehension.

Similar observations also apply to the comparison with the refactoring approach by Binkley et al. [1]. Their emphasis is on full automation, and they offer an Eclipse plugin for conducting six elementary refactorings. They focus on our fourth step only, and assume aspect mining has resulted in `@begin-aspect` and `@end-aspect` annotations in the code. As an example, one of their six refactorings moves individual calls to separate aspects, after which a (non-trivial) pointcut abstraction step is needed to merge the results. Our approach eliminates the need for this complex abstraction step, thanks to the sort-based integration between aspect mining and refactoring (refactoring is based on a full concern model in our case). Like us, they use JHOTDRAW as one of their case studies. Somewhat surprisingly, they do not report any of the limitations that we identified, although their results exhibit the same limitations.

## 8. Concluding remarks

In this paper, we proposed an integrated strategy for migrating crosscutting concerns to aspect-oriented programming. We presented in detail the refactoring step of our strategy, and applied the entire migration process to concerns in an open-source application. Furthermore, we discussed the challenges of refactoring crosscutting concerns to aspects and how these could impact the design and implementation of automatic aspect refactoring.

The contributions of this work can be summarized as:

- An integrated strategy for migrating crosscutting concerns to AOP solutions;
- An aspect refactoring approach based on crosscutting concern sorts and a set of refactoring templates;
- A report on our experience with migrating concerns in a real system to aspects and the challenges of this process. This report is useful for assessing the present support for refactoring and the feasibility of automatic aspect refactoring for various categories (that is, sorts) of crosscutting concerns.
- AJHOTDRAW, a show-case for aspect refactoring in an open-source implementation that can be further used by researchers and practitioners to evaluate aspect-based solutions to crosscutting concerns.

AJHOTDRAW provides a code base for related research to measure code improvements due to aspect code. Furthermore, this work provides us with the hands-on experience for designing and implementing sort-based aspect refactoring. We plan to extend our tool support for concern documentation, SOQUET, with aspect refactoring options. The refactoring would apply to each query documenting a sort instance, and hence benefit from the description of the concerns available by the query results. We appreciate that a significant effort would go into the design and implementation of wizards to deal with the various reported challenges.

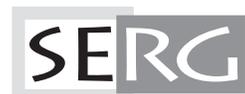